\documentclass[12pt]{article}
\usepackage{latexsym,amssymb,amsmath} 
\usepackage{graphicx}
\usepackage{epsfig}
\usepackage{verbatim}

\usepackage{amsfonts}


\title{\textbf{Tailored photon pair preparation relying on full group velocity matching in fibre-based spontaneous four wave mixing}}

\author{Karina Garay-Palmett$^1$\thanks{Corresponding author. Email: kgaray@cicese.mx}, Ra\'ul Rangel-Rojo$^1$ and Alfred B. U'Ren$^{1,2}$
\\\vspace{1pt}\small\em{ $^1$Departamento de \'{O}ptica, Centro de
Investigaci\'{o}n Cient\'{i}fica y Educaci\'{o}n Superior de
Ensenada}
\\\vspace{1pt}\small\em{(CICESE), Baja
California, 22860, M\'exico}
\\\vspace{1pt}\small\em{$^2$Instituto de Ciencias Nucleares, Universidad Nacional Aut\'onoma
de M\'exico,}
\\\vspace{1pt}\small\em{ apdo. postal 70-543, M\'exico 04510 DF}}

\begin{document}
\DeclareGraphicsExtensions{.eps} 

\maketitle
\begin{abstract}

We study photon pair generation through scalar spontaneous four-wave
mixing in single-mode fibre and for frequency-degenerate pumps; we
concentrate on source geometries which fulfil full group velocity
matching (GVM), {\em i.e.} where the pump, signal and idler
propagate at identical group velocities.  We discuss two
experimental techniques which permit the attainment of  full GVM, and
discuss the resulting two-photon state properties.  In particular, we
show that  full GVM can lead to sources which approach phasematching
unconstrained by dispersion and therefore with a remarkably large
bandwidth.  We also discuss the generation of nearly-factorable
states as an application of  full GVM.\\


\end{abstract}

\section{Introduction}

The process of spontaneous four wave mixing (SFWM) in optical fibre,
in which two pump photons are jointly annihilated to generate a
signal and idler photon pair, leads to a remarkably versatile photon
pair source\cite{fiorentino02,rarity05,fan05}.  Indeed, careful
choice of the properties of the pump fields, which can be spectrally
degenerate or non-degenerate, and of the properties of the fibre,
can lead to two-photon states with widely disparate characteristics,
ranging from factorable to highly entangled.  Let us note that for a
photon-pair source where the generation process is constrained to a
single transverse mode, as in single-mode optical fibre, the
resulting photon pairs are unentangled in transverse wavevector,
leaving frequency as the only continuous-variable degree of freedom
where entanglement may reside. While spectrally factorable
two-photon states are essential for heralding of pure single
photons\cite{uren05},  at the opposite extreme highly entangled
photon pairs\cite{zhang07,odonnell07} have useful applications in
quantum-enhanced two-photon absorption\cite{dayan05}, single-photon
wavepacket teleportation\cite{molotkov98}, and quantum optical
coherence tomography\cite{nasr03}, among others.  It has been shown
that both factorable\cite{grice01,mosley08} and highly
entangled\cite{zhang07,odonnell07} photon pairs may be generated
through parametric downconversion (PDC) in second-order non linear
crystals. However, SFWM greatly facilitates (with respect to
PDC) the fulfilment of the conditions required for spectrally
engineered states including both factorable and highly entangled
ones. Furthermore, the guided nature of a spontaneous non-linear process
in fibre leads to a higher parametric gains and evidently suppresses
losses due to fibre-coupling. Thus, for example, as shown in
Ref.\cite{garay07}, any fibre with two zero dispersion frequencies
permits the generation of two-photon states with an arbitrary
orientation in the joint frequency space $\{\omega_s,\omega_i\}$,
including factorable states as a special case.

In this paper, we study SFWM sources in the degenerate pumps regime
for which full group velocity matching is fulfilled, {\em i.e.} where the pump, signal and idler propagate at identical group velocities. We show that a source with these characteristics may
approach phasematching unconstrained by dispersion, and therefore
lead to a remarkably large phasematching bandwidth, both for the
pump and the generated light.  Likewise, we show that long fibres
which fulfil a certain condition on the second-order dispersion
coefficients, permit the generation of nearly factorable states, in
a source geometry which is well suited for multiple-pair generation
through a large parametric gain.

\section{Theory}

The quantum state of photon pairs produced by spontaneous, scalar
four wave mixing in an optical fibre of length $L$ can be obtained
following a standard perturbative approach.  It is given by

\begin{align}
\label{eq: state} &  |\Psi\rangle= |0\rangle_{s}|0\rangle_{i}
+\kappa\int d\omega_{s}\int d\omega_{i} F\left(
\omega_{s},\omega_{i}\right)  \left| \omega_{s}\right\rangle _{s}
\left|  \omega_{i}\right\rangle _{i}.
\end{align}

\noindent Here, $\kappa$ represents the generation efficiency and
$F\left(\omega_s,\omega_i\right)$ is the joint spectral amplitude
function (JSA), which describes the spectral entanglement properties
of the photon pairs. For frequency-degenerate pumps, $F \left(
\omega_{s},\omega_{i}\right)$ can be expressed as\cite{garay07}

\begin{align}
\label{eq: JSA}F\left(  \omega_{s},\omega_{i}\right)   &  = \int
d\omega^{\prime}\alpha\left(  \omega^{\prime}\right)  \alpha
\left(
\omega_{s}+\omega_{i}-\omega^{\prime}\right) \nonumber\\
&  \times\mbox{sinc}\left[L\Delta k\left(  \omega^{\prime}%
,\omega_{s},\omega_{i}\right) /2 \right]  \mbox{exp} \left[ iL\Delta k\left(
\omega^{\prime},\omega_{s},\omega_{i}\right)/2 \right].
\end{align}

The JSA function is given in terms of the phase mismatch function $
\Delta k\left( \omega _{1},\omega _{s},\omega _{i}\right) =k\left(
\omega _{1}\right) +k\left( \omega _{s}+\omega _{i}-\omega
_{1}\right) -k\left( \omega _{s}\right) -k\left( \omega _{i}\right)
-2  \gamma P$, which includes a self/cross-phase modulation
contribution for the pump with peak power $P$, characterized by the
nonlinear parameter $\gamma$, and the spectral shape
$\alpha(\omega)$ of the pump.

It can be shown that in the case where the pumps are narrowband, it
becomes possible to obtain an analytic expression in closed form for
the JSA

\begin{align}
\label{JSA1}
F_{cw}(\omega_s,\omega_i) =N \delta(\omega_s+\omega_i-2\omega_{p})
\mbox{sinc}[L\Delta k_{cw}(\omega_s,\omega_i)/2]\exp[i L\Delta k_{cw}(\omega_s,\omega_i)/2],
\end{align}

\noindent in terms of a normalization constant $N$ and the monochromatic pump phasemismatch $\Delta k_{cw}$

\begin{equation}
\Delta
k_{cw}(\omega_s,\omega_i)=2 k\left[\left(\omega_s+\omega_i\right)/2\right]
-k(\omega_s)-k(\omega_i)-2 \gamma P.
\end{equation}

For broadband, degenerate pumps it is likewise possible to obtain an
expression for the JSA in closed analytic form, if we approximate
the phasemismatch $L \Delta k$ by a truncated power series in
$\omega$, and model the pump to have a Gaussian spectral envelope,
i.e. $S(\omega)=\exp[-(\omega-\omega_p)^2/\sigma^2]$ (with central
frequency $\omega_p$ and bandwidth $\sigma$). Keeping up to
second-order terms in the power series, it can be shown that the JSA
may be expressed as $F(\nu_s,\nu_i)=N'
\alpha(\nu_s,\nu_i)\phi(\nu_s,\nu_i)$, where we have defined
frequency detunings $\nu_\mu=\omega_\mu-\omega_{\mu 0}$, with
$\mu=s,i$ and where $\omega_{\mu 0}$ represent central phasematched
frequencies. Here, $N'$ represents a normalization constant,
$\alpha(\nu_s,\nu_i)=\exp[-(\nu_s+\nu_i)^2/(2 \sigma^2)]$ and
$\phi(\nu_s,\nu_i)=\Phi(C_0;Z(\nu_s,\nu_i))$ with

\begin{equation}
\label{phi} \Phi(a;x)=\exp(-x^2) \left[ \mbox{erf}(ix\sqrt{1-i
a})-\mbox{erf}(ix) \right]/(a x),
\end{equation}

\noindent where $\mbox{erf}(\cdot)$ denotes the error function, $C_0=\tau_p^{(2)} \sigma^2/2$, and

\begin{equation}
\label{zeta}
Z(\nu_s,\nu_i)=  \sqrt{(\nu_s+\nu_i)^2 -4\beta(\nu_s,\nu_i)/\tau_p^{(2)}}/(\sqrt{2}\sigma),
\end{equation}

\noindent in terms of the phasemismatch product $L\Delta k$ evaluated within our second-order approximation, $\beta(\nu_s,\nu_i)$,

\begin{eqnarray}
\label{delt}\beta( \nu_{s},\nu_{i})=L\Delta
k^{(0)}+\tau_s^{(1)}\nu_s+\tau_i^{(1)}\nu_i+\tau_s^{(2)}\nu_s^2+\tau_i^{(2)}\nu_i^2+
\tau_p^{(2)}\nu_s\nu_i.
\end{eqnarray}

In the previous equations we have introduced the definitions
$\tau^{(n)}_\mu=L[k^{(n)}(\omega_p)-k^{(n)}(\omega_{\mu 0})]/n!$,
with $\mu=s,i$, and   $\tau^{(2)}_p=L k^{(2)}(\omega_p)$, in terms
of the $n$th order derivatives  $k^{(n)}\left( \omega_{\mu}\right)
=d^{n}k/d\omega^{n} |_{\omega=\omega_{\mu o}}$.

We are particularly interested in cases where the three fields
(pump, signal and idler) propagate along the fibre with identical
group velocities, which we refer to as full group velocity matching.
In this case, the $\tau^{(1)}_{\mu}$ terms in Eq.(\ref{delt})
vanish. Assuming that phasematching is fulfilled at frequencies
$\omega_{s 0}$ and $\omega_{i 0}$, the $\Delta k^{(0)}$ term (see
Eq.(\ref{delt})) also vanishes.  In this case, the two-photon state
properties are fully determined to lowest order by group velocity
dispersion terms.  Let us note that for frequencies sufficiently
removed from $\omega_{s 0}$ and $\omega_{i 0}$, Eqns.(\ref{phi}) and
(\ref{zeta}) are no longer valid.    This description is likewise not
valid if  $\tau_p^{(2)}=0$, or if all second order dispersion
coefficients vanish.

\section{Phasematching properties of optical fibres with more than one zero dispersion frequency}

\begin{figure}[ht]
\begin{center}
\begin{minipage}{140mm}
\begin{center}
\resizebox*{11cm}{!}{\includegraphics{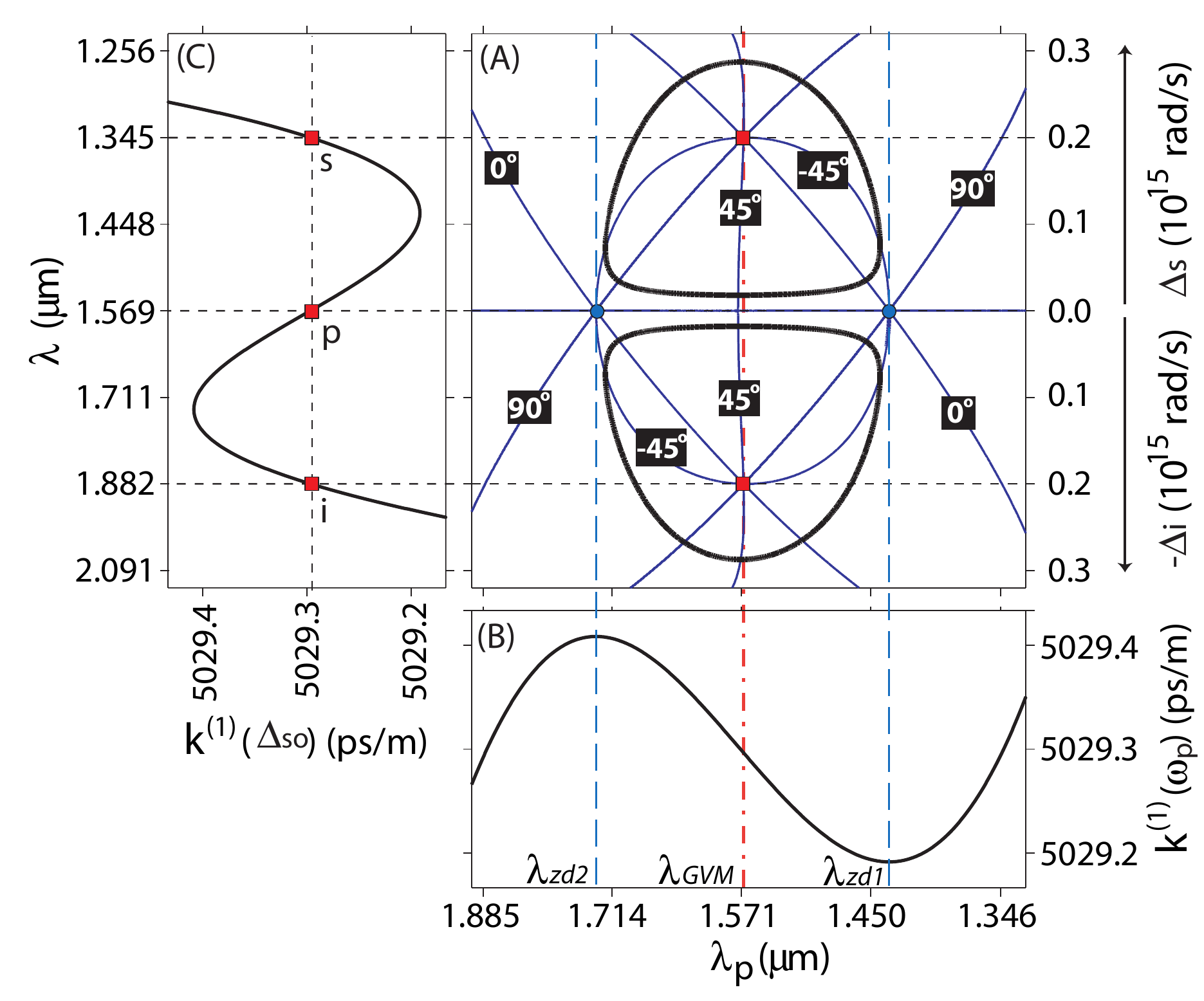}}%
\end{center}
\caption{Phase-matching diagram for a SIF with core radius
$r=1.652\,\mu$m and index contrast $\Delta_n=0.0274$. (A)
Thick black line: $\Delta k=0$ contour; Thin blue lines: contours defined by constant orientation angle.
(B) $k^{(1)}$ vs $\omega$. (C)
$k^{(1)}$ vs $\Delta_{s0}=\omega-\omega_{GVM}$.}%
\label{figura1}
\end{minipage}
\end{center}
\end{figure}

In this paper we consider the use of step-index fibres (SIF), in
both, the low and high dielectric contrast regimes, depending on the
application. We are particularly interested in SFWM sources based on
fibres which exhibit two zero dispersion frequencies (ZDF) in the
spectral range of interest.  Note that while some of the geometries
to be considered exhibit a third ZDF, if this third ZDF is
sufficiently removed from the two ZDFs under consideration, its
effects are limited. As was shown in Ref.\cite{garay07}, fibres with
two ZDFs permit a remarkable flexibility in terms of engineering the
spectral entanglement properties of the resulting two-photon states.
One way to represent the phasematching (PM) properties of a given
fibre is to plot on a generated frequencies vs pump frequencies
diagram, pairs of frequencies which yield perfect PM, {\em
i.e.} $\Delta k=0$. Such a plot is presented in
Fig.~\ref{figura1}(A) by the thick black contour, for
a SIF with a fused silica cladding and a core such that the index
contrast is $\Delta_n=(n_{co}-n_{cl})/n_{co}=0.0274$
($n_{co}$/$n_{cl}$ are the core/cladding indices of refraction),
with a core radius of $r=1.652 \,\mu$m.  In this diagram, the
horizontal axis corresponds to the pump frequency, and the vertical
axis corresponds to the generated frequencies, expressed as a
detuning $\Delta$ from the pump frequency, {\em i.e.}
$\Delta=\omega-\omega_p$ (the top/bottom half of the diagram
corresponds to the signal/idler photon).  It can be seen from this
illustration that parametric generation occurs for a range of pump
frequencies essentially corresponding to the interval between the
two ZDFs; for the specific case shown, the zero dispersion
wavelengths are $\lambda_{zd1}=2 \pi c/\omega_{zd1}=1434.3\,$nm, and
$\lambda_{zd2}=2 \pi c/\omega_{zd2}=1733.5\,$nm (this fibre has a
third zero dispersion wavelength $\lambda_{zd3}=2 \pi
c/\omega_{zd3}=2224.3\,$nm). The ZDFs  are marked on the
diagram by two blue vertical long-dashed lines;
Fig.~\ref{figura1}(B) shows that these frequencies correspond to the
two points with vanishing slope on a $k^{(1)}$ vs $\omega$ plot (and
therefore with $k^{(2)}=0$). Fig.~\ref{figura1}(B) illustrates an
important property of fibres with two ZDFs: within the frequency interval bounded by the two ZDFs, anomalous group velocity dispersion is observed, which is intimately connected with the appearance of
closed-loop PM contours, as in Fig.~\ref{figura1}(A). Note
that for a finite pump power, self/cross phase modulation suppresses phase
matching for degenerate signal/idler frequencies. As a consequence, modulation
instability splits the trivial solution into two closely-spaced branches,
where the separation increases with pump power\cite{agrawal2007};
here we have assumed $\gamma=70\,\mbox{W}^{-1} \mbox{km}^{-1}$ and
$P=2\,W$.

%

The two-photon spectral characteristics are governed to first order
by the coefficients $\tau_s^{(1)}$ and $\tau_i^{(1)}$.  It was shown
in Ref.~\cite{garay07} that, for a broadband pump, specific
conditions on these parameters, which quantify the group velocity
mismatch between the pump pulse and the generated photons, permit
the generation of two photon states with useful properties including
the important class of factorable states.  It is instructive to plot
in the $\{\omega_p,\Delta\}$ space of Fig.~\ref{figura1}(A) the GVM
contours defined by the conditions $\tau_s^{(1)}=0$, equivalent to
$k^{(1)}_s=k^{(1)}_p$ (GVM between the pump and the signal photon),
and $\tau_i^{(1)}=0$, equivalent to $k^{(1)}_i=k^{(1)}_p$ (GVM
between the pump and the idler photon).  In
Fig.~\ref{figura1}(A) these contours correspond to the thin blue
lines and are labelled by
$\theta_{pm}=-\arctan(\tau_s^{(1)}/\tau_i^{(1)})$, which was shown
in Ref.~\cite{garay07} to correspond to the orientation angle of the
phasematched region in the joint frequency space
$\{\omega_s,\omega_i\}$;  the two GVM scenarios above lead to values
$\theta_{pm}=0^\circ$ and $\theta_{pm}=90^\circ$, respectively.  In
the figure we have in addition included the contours corresponding
to $\theta_{pm}=\pm 45^\circ$.  Pairs of values
$\{\omega_p,\Delta\}$  for which the GVM contours for all
$\theta_{pm}$ intersect, to be referred to as FGVM points, indicate
$\{\omega_p,\Delta\}$ values for which \textit{full group velocity
matching} is attained, {\em i.e.} $k^{(1)}_p=k^{(1)}_s=k^{(1)}_i$.
In general, there are four such points; two of these are
$\{\omega_{zd1},0\}$ and $\{\omega_{zd2},0\}$, indicated with
blue circle markers in Fig.~\ref{figura1}(A), (i.e.
identical pump, signal and idler frequencies, with either
$\omega_p=\omega_{zd1}$ or $\omega_p=\omega_{zd2}$). These two
points are not phasematched, except in the limit of zero pump power.
The other two FGVM points are within each of the two PM loops, as
illustrated with red square markers in
Fig.~\ref{figura1}(A), and are therefore typically not phasematched.
We will refer to the pump frequency at which this second set of FGVM
points occur as $\omega_{GVM}$; in Fig.~\ref{figura1}(A)
this frequency is marked by the red dash-dot line. It is shown in
Fig.~\ref{figura1}(C), which shows a plot of $k^{(1)}$ vs
$\Delta_{s0}=\omega-\omega_{GVM}$, that the signal and idler
frequencies corresponding to $\omega_p=\omega_{GVM}$ indeed lead to
$k^{(1)}_s=k^{(1)}_i=k^{(1)}_p$. In what follows, we study methods
for achieving phasematched full GVM, and discuss the result of this
on the two-photon state.

\begin{figure}
\begin{center}
\begin{minipage}{140mm}
\begin{center}
\resizebox*{12cm}{!}{\includegraphics{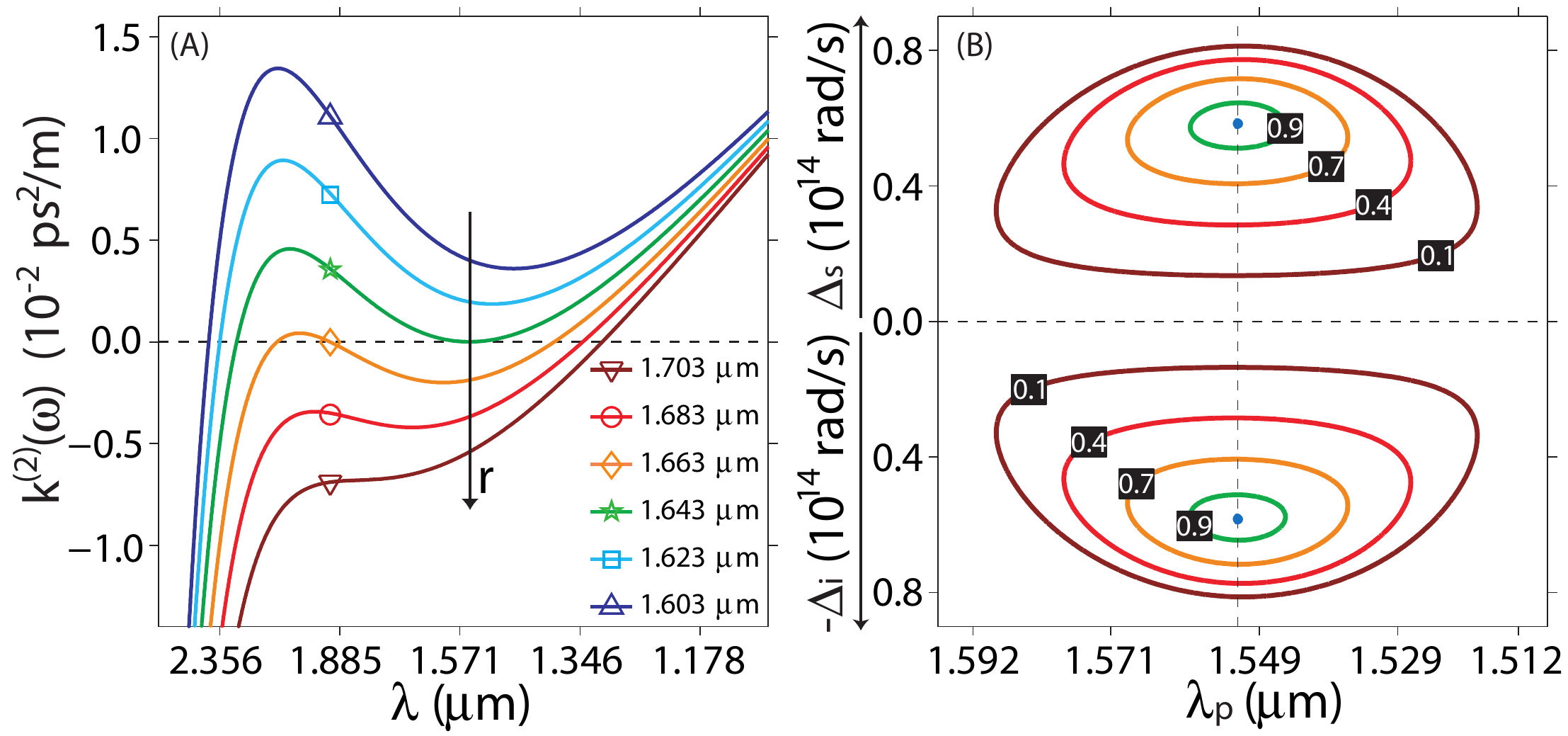}}%
\end{center}
\caption{(A) $k^{(2)}$ vs $\omega$ for different core radii $r$; we
have assumed a fused-silica step-index fibre with $\Delta_n=0.0274$.
(B) Phase-matching contours for different pump power levels
(assuming $\Delta_n=0.0274$ and $r=1.644\,\mu$m, power in W
indicated within black squares); for $P=0.95\,$W the PM loops become
point-like.}%
\label{figura2}
\end{minipage}
\end{center}
\end{figure}

Let us concentrate on the two FGVM points at $\omega_p=\omega_{GVM}$
(see Fig.~\ref{figura1}).  For most source geometries, these points
are not phasematched, since they occur within the PM loops.  Here,
we will explore two methods which lead to phasematching at these
points.   As we will see, in a realistic source design these two
methods are most effective when used in conjunction, though each one
is in principle independently capable of yielding photon pairs
characterized by full GVM.  Two of the FGVM points are directly
linked to the ZDFs, which as is well known exhibit a strong
dependence on the fibre core radius.   In fact, the location of all
four FGVM points can be varied through changes in the fibre core
radius.  In general terms, if the waveguide contribution to the
dispersion is sufficiently strong relative to the material
contribution ({\em i.e.} if there is a strong nucleus-cladding
dielectric contrast), then more than one ZDF may exist, two of which
may approach each other for certain core radii. This behavior is
illustrated in Fig.~\ref{figura2}(A), which shows for a SIF with a
fused silica cladding and a core such that the index contrast is $\Delta_n=(n_{co}-n_{cl})/n_{co}=0.0274$ ($n_{co}$/$n_{cl}$ are the core/cladding indices of refraction), $k^{(2)}$ plotted as a
function of $\omega$ for different core radii $r$.  For
$1.643\,\mu$m$<r<1.665\,\mu$m, three ZDFs exist, where the two
higher ones become degenerate at $r=1.643\,\mu$m.  For the core
radius at which the two higher ZDFs merge, the resulting degenerate
ZDF fulfils $k^{(2)}=k^{(3)}=0$.   Also, in this case the four FGVM
points merge into a single point.  Note that a further reduction of
the core radius leads to the suppression of PM.  A fibre with two
degenerate ZDFs at $\omega_{zd}$, pumped at $\omega_{zd}$, leads to
the generation of photon pairs characterized by full GVM, centered
also at $\omega_{zd}$.

Alternatively, it is possible to exploit the self/cross phase
modulation contribution to achieve PM at the two
$\omega_p=\omega_{GVM}$ FGVM points.  As has already been mentioned,
the shape of the PM loops exhibits a dependence on the pump power
$P$. At small power levels, this effect manifests itself as the
appearance of symmetric sidebands, detuned by $\pm (2\gamma
P/|k^{(2)}|)^{1/2}$ from the pump frequency\cite{agrawal2007}; in
contrast, the effect on the outer portions of the PM loops tends to
be comparatively weaker. For example, in Fig.~\ref{figura1}(A), the
effect of self/cross phase modulation is only apparent in the inner
loop. Thus, for small power levels, the PM loop splits into two
separate loops.  However, for sufficiently large pump power levels,
the entire PM loops, including the outer areas, shrink.  At a
specific power level, the two PM loops become each a single point,
which in fact corresponds to each of the two $\omega_p=\omega_{GVM}$
FGVM points.  Note that increasing the power further beyond this
value suppresses PM.  This is illustrated in Fig.~\ref{figura2}(B),
where we show for a fibre with index contrast $\Delta_n=0.0274$, and
core radius $r=1.644\,\mu$m the PM loops resulting for five
different pump power levels ($P_1=0.10\,$W, $P_2=0.40\,$W,
$P_3=0.70\,$W, $P_4=0.90\,$W, $P_5=0.95\,$W). Power level $P=P_5$
corresponds to that for which the PM loops become point-like. Thus,
a pump centered at $\omega_{GVM}$ with the specific pump power
leading to point-like PM loops, produces photon pairs characterized
by full GVM.

\begin{figure}[ht]
\begin{center}
\begin{minipage}{140mm}
\begin{center}
\resizebox*{12cm}{!}{\includegraphics{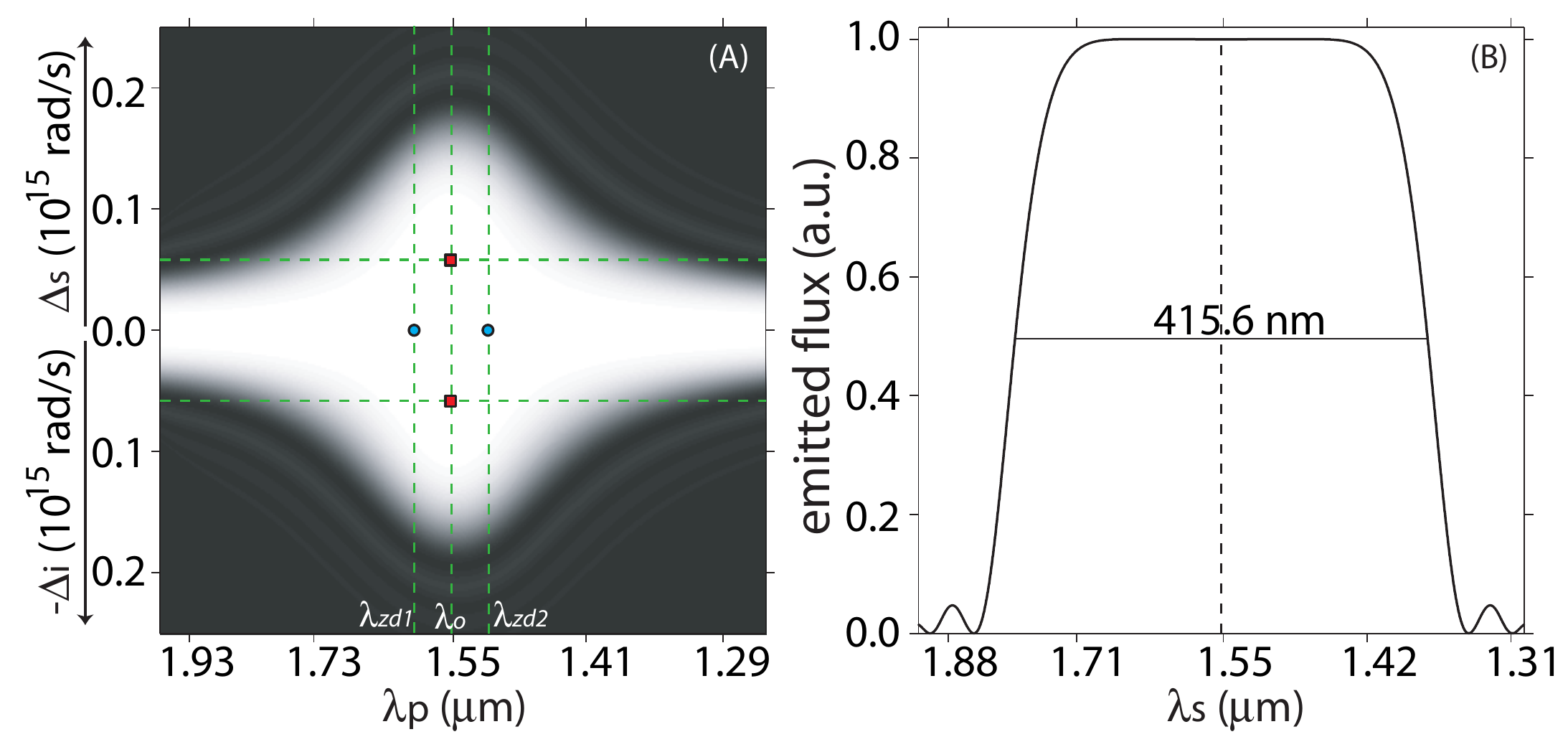}}%
\end{center}
\caption{(A) PM function for a step-index fiber of length $L=50\,$cm
(with the other parameters as in Fig.~\ref{figura2}(B)). Perfect PM
is achieved in the two FGVM points which occur at $\lambda_{GVM}=2
\pi
c/\omega_{GVM}=1552.1\,$nm (red square markers); the blue circle markers indicate the other two FGVM points. (B) SFWM singles spectrum for $\omega_p=\omega_{GVM}$.}   %
\label{figura3}
\end{minipage}
\end{center}
\end{figure}

What are the properties of photon pairs characterized by full GVM?
On the one hand, as is clear from the discussion above, PM is
achieved at individual points (in $\{\omega_p,\Delta\}$ space).   On
the other hand, because the normally-dominant first order terms in
the phasemismatch (see Eq.(\ref{delt})) are suppressed, $\Delta k$
grows only weakly, as governed by second order terms, away from the
points where perfect PM is achieved (and which correspond to FGVM
points). This can translate into a large area in generated vs pump
frequencies space where essentially perfect phasematching is
attained.   Thus, for a given pump frequency, broadband SFWM is
generated, and in addition the pump frequency may be varied over a
large spectral interval maintaining broadband SFWM photon-pair
generation.  Fig.~\ref{figura3} illustrates this behavior, for a
fused-silica SIF of length $L=50\,$cm with an index-contrast of
$\Delta_n=0.0274$.  Fig.~\ref{figura3}(A) shows a phasematching
diagram, plotted in $\{\omega_p,\Delta\}$ space, where for a
monochromatic pump centered at $\omega_p$ (horizontal axis) we have
computed the resulting  singles spectrum $S(\omega)=\mbox{sinc}^2[L
\Delta k_{cw}(\omega,2 \omega_p-\omega)/2]$.  We have selected a
fibre radius, $r=1.644\,\mu$m (as in Fig.~\ref{figura2}(B)), for
which (see Fig.~\ref{figura2}(A)), the two ZDFs are nearly
coincident, and for which the PM loops shrink down to the FGVM
points through the self/cross phase modulation contribution. Thus,
we have relied on a mixture of the two mechanisms above to obtain
phasematched FGVM. On the one hand, the first mechanism discussed
(based on choosing the core radius) leads to the merging of the four
FGVM points, and a realistic pump power level suppresses PM. On the
other hand, the second mechanism discussed (based on choosing the
pump power level) tends to require large power levels to close down
the loops. By making the loops small (for nearly coincident ZDFs),
only a relatively small power level is required for the PM loops to
become single points.  This is the situation presented in
Fig.~\ref{figura3}(A), where we have assumed a pump power of
$P=0.9496\,$W. The experimental implementation of these PM
techniques necessitates an optical fibre with suitable dispersion properties. In the case of photonic crystal
fibre and SIF with the appropriate core-cladding dielectric
contrast, the required core radius can be reliably obtained through existing tapering techniques\cite{wadsworth02,Brambilla05,Foster08}.

In Fig.~\ref{figura3}(A) we have indicated the location of the four
FGVM points.  Perfect PM is achieved in the two
$\omega_p=\omega_{GVM}$ FGVM points (red square markers in the
figure). Note that the PM bandwidth is remarkably broad; the
phasematched areas around these two points merge into a single,
broad region, shown in white, characterized by nearly-perfect PM.
Fig.~\ref{figura3}(B) shows, for $\omega_p=\omega_{GVM}$, (with
$\lambda_{GVM}=2 \pi c/\omega_{GVM}=1552.1\,$nm) the resulting SFWM
singles spectrum, with a full width at half maximum bandwidth of
$\Delta \lambda_{max}=415.6\,$nm. An indication of the large pump
bandwidth is that the pump wavelength may be varied between $1436.0\,$nm
and $1750.0\,$nm, maintaining a generated SFWM bandwidth of $\Delta
\lambda \ge \Delta \lambda_{max}/2$.   A possible application for a
source with these characteristics, is a broadband parametric
amplifier\cite{radic03}. Indeed, with a phasematched region such as
that shown in Fig.~\ref{figura3}, the central frequencies for both
the pump $\omega_p$ and the seed $\omega_s$, to be amplified, could
be chosen so that $\{\omega_p,\omega_s\}$ is anywhere within the
white area. Such an amplifier would exhibit a constant level of
noise within the phasematched area resulting from SFWM, {\em i.e.}
without spectral structure.  In addition, full GVM leads to the
important property that the amplified seed does not develop
additional spectral structure\cite{brainis05}. Note that in
Ref.\cite{garay08} we have presented an alternative recipe for
ultrabroadand SFWM generation based on controlling higher-order
dispersive terms.  This technique (from Ref.\cite{garay08}) permits
an even wider generation bandwidth as compared to the technique
presented here at the cost, however, of a greatly reduced pump
bandwidth, {\em i.e.} for the technique from Ref.\cite{garay08}, the
PM region in a diagram similar to Fig.~\ref{figura3}(A) would be
wider vertically, and much narrower horizontally.

\section{Generation of nearly-factorable photon pairs}

\begin{figure}[ht]
\begin{center}
\begin{minipage}{140mm}
\begin{center}
\resizebox*{12cm}{!}{\includegraphics{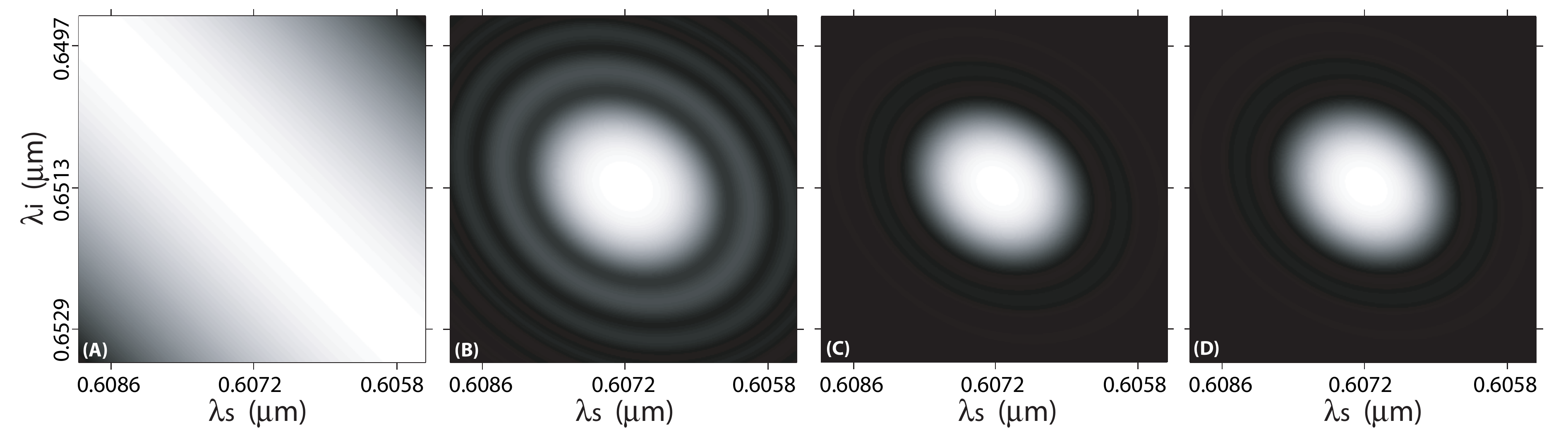}}
\end{center}%
\caption{Joint spectral intensity (JSI) for an air-guided bismuth
borate glass rod, for which PM is attained at the two
$\omega_p=\omega_{GVM}$ FGVM points.  (A) Pump envelope function
$\alpha(\nu_{s},\nu_{i})$. (B) Phase-matching function $\phi\left(
\nu_{s},\nu_{i}\right)$. (C) Analytic JSI, from Eqns. (\ref{phi} to
\ref{delt}). (D) JSI obtained by numerical integration of Eq.
(\ref{eq: JSA}).}%
\label{figura4}
\end{minipage}
\end{center}
\end{figure}

As has already been pointed out, for FWM sources exhibiting full
GVM, while perfect PM occurs at isolated points (e.g. in
$\{\omega_p,\Delta\}$ space), $\Delta k$ increases only slowly away
from these points.   In the previous section we exploited this weak
spectral dependence of $\Delta k$ to design a source which permits a
remarkably broad generation bandwidth, over a wide range of pump
frequencies. In this section, we exploit the fact that perfect PM
occurs at isolated points to design relatively narrowband photon
pair sources, where in addition by imposing a certain condition on
the second order dispersion coefficients it becomes possible to
approach factorability.   For sufficiently narrowband photon pairs,
the description in terms of Eqns.(\ref{phi}) and (\ref{zeta})
becomes valid. Note that $Z(\nu_s,\nu_i)$  (see Eq.(\ref{zeta})) is
independent of the fibre length $L$; the only dependence on $L$ is
through parameter $C_0$.  The effect of increasing $C_0$ (which
depends linearly on $L$) is to decrease the spectral width of the PM
function.  Thus, for sufficiently long fibres, the phasematched
region in $\{\omega_p,\Delta\}$ space becomes localized around the
points which yield perfect PM.  If the core radius and power level are
chosen so that perfect PM occurs at the two $\omega_p=\omega_{GVM}$
FGVM points, then on the one hand $\Delta k^{(0)}=0$ and on the
other hand $\tau_s^{(1)}=\tau_i^{(1)}=0$.  Under these conditions,
from Eqns.(\ref{zeta}) and (\ref{delt}) we can show that the
phasematched regions become circular (i.e. the contours
$Z(\nu_s,\nu_i)=$constant become circles) if the condition
$\tau_s^{(2)}=\tau_i^{(2)} \gg \tau_p^{(2)}$ is satisfied.

If this condition is satisfied, then for a sufficiently broad pump
bandwidth (so that the joint amplitude is determined solely by
function $\phi(\omega_s,\omega_i)$; see text before Eq.(\ref{phi}),
the two photon state becomes factorable. The technique outlined here
for the generation of factorable photon pairs presents some
significant advantages over the techniques presented in
Ref.~\cite{garay07}. Indeed, an important practical difference is
that in the techniques presented earlier, it is essential to select
 a particular fibre length for a given pump bandwidth.  Thus, increasing the fibre
length, e.g. in order to increase the source brightness, would lead
to the need for a simultaneous reduction of the pump bandwidth so as
to maintain factorability.   In the technique presented here,
increasing the fibre length (while maintaining the pump bandwidth
constant) has the effect of reducing the generation bandwidth, while
maintaining factorability. The resulting decoupling of fibre length
and pump bandwidth significantly enhances the experimental
flexibility of this type of source.  Furthermore, note that it
becomes possible to obtain increasingly narrowband and high-flux
nearly-factorable photon pair generation (in principle without
limit) by increasing the fibre length.   Indeed, we believe that
this source could be well suited for experiments intended to
generate multiple pairs simultaneously, through a large parametric
gain\cite{rohde}.

The fulfilment of condition $\tau_s^{(2)}=\tau_i^{(2)} \gg
\tau_p^{(2)}$ is facilitated if $k^{(2)}$ exhibits a strong
dependence on frequency.  This translates into the need for
relatively large higher-order dispersive terms, which is made
possible by a fibre with a strong waveguide dispersion, i.e. with a
strong nucleus-cladding dielectric
contrast\cite{Tong06,Ebendorff04,Gopinath}.  In
Fig.~\ref{figura4} we illustrate the design of a nearly-factorable
photon pair source based on these ideas.   We have assumed a fibre
composed of an air-guided bismuth borate glass rod with radius
$r=0.205\,\mu$m; the combination of a particularly high-index glass
and air guiding leads to a remarkably large dielectric contrast. The
fibre radius is chosen so that the two ZDFs are in relative
proximity, with $\lambda_{zd1}=2 \pi c/\omega_{zd1}=616.2\,$nm and
$\lambda_{zd2}=2 \pi c/\omega_{zd2}=641.7\,$nm. The pump frequency
must coincide with $\omega_{GVM}$ (with $\lambda_{GVM}=2 \pi
c/\omega_{GVM}=628.5\,$nm), while the FWHM pump bandwidth, $\Delta
\lambda=6.29\,$nm, is large enough that the two-photon state is
fully determined by the function $\phi(\omega_s,\omega_i)$. The
fibre length, $L=100\,$m, is chosen so as to restrict the emission
bandwidth (to around $1.2\,$nm FWHM). Finally, the pump power,
$P=9.75\,$W, is chosen so that the PM loops become point-like (at
$\lambda_{s0}=2 \pi c/\omega_{s0}=607.2\,$nm and $\lambda_{i0}=2 \pi
c/\omega_{i0}=651.3\,$nm) [see also Fig.~\ref{figura2}(B)]; we have
assumed a nonlinear coefficient of $\gamma=550
\,\mbox{W}^{-1}\mbox{km}^{-1}$. Figure~\ref{figura4} shows, plotted
as a function of the signal and idler frequencies: the pump envelope
function $\alpha(\omega_s,\omega_i)$ in panel (A), the phasematching
function $|\phi(\omega_s,\omega_i)|$ in panel (B),  the joint
spectral intensity (JSI)
$|\alpha(\omega_s,\omega_i)\phi(\omega_s,\omega_i)|^2$ in panel (C),
and the JSI obtained through numerical integration of Eq.(\ref{eq:
JSA}) in panel (D).  It is evident from the figure that the agreement
between the analytic and numerical calculations is excellent.
Figs.~\ref{figura4}(C) and (D) show that the resulting two-photon
state is essentially factorable; the detection of an idler photon
would herald a single photon in the signal mode with purity
$P=\mbox{Tr}(\rho_s^2)=0.88$ (where $\rho_s$ is the reduced density
operator of the signal mode).

\section{Conclusions}

We have analyzed the spectral properties of two-photon states
produced by spontaneous four wave mixing (SFWM) in a single mode optical
fibre and in the degenerate-pumps regime, focusing our attention on
fibres with more than one zero dispersion frequency (ZDF).   We have
explored the conditions under which we expect SFWM photon pair
generation characterized by full group velocity matching (GVM), {\em
i.e.} where the pump, signal and idler propagate at identical group
velocities.  We have presented two routes to phasematched full GVM,
one based on engineering the fibre to have coincident ZDFs, and the other based on the self/cross phase
modulation term; these two methods are most effectively used
together.  We have explored the consequences of full GVM, when satisfied together with phasematching (PM),  on the two-photon
spectral properties . We have shown that for a source with these
characteristics, while perfect PM occurs at isolated pump
and generated frequencies, because the phasemismatch grows only
slowly from these points, we obtain a very large pump and
signal/idler bandwidth over which essentially perfect PM
is obeyed.   Thus, it becomes possible to generate broadband SFWM
photon pairs, for a large range of pump frequencies.  Likewise, we have shown that, using long fibers which satisfy the condition $\tau_s^{(2)}=\tau_i^{(2)} \gg
\tau_p^{(2)}$ on the second-order dispersion coefficients, we can exploit the fact that
perfect PM occurs at isolated frequencies to obtain relatively
narrowband, nearly-factorable two-photon states.  We expect that these
results will be useful for the practical implementation of
fibre-based photon pair sources for quantum information processing
applications.

\section*{Acknowledgement(s)}
KGP, and RRR acknowledge support from CONACYT
through project 46492; ABU acknowledges support from
CONACYT through project 46370.


\begin{thebibliography}{99}

\bibitem[1]{fiorentino02} Fiorentino, M.; Voss, P. L.; Sharping, J. E.; Kumar, P. {\em IEEE Photon. Technol. Lett.} {\bf 2002} {\em 14}, 983-985. 

\bibitem[2]{rarity05} Rarity, J. G.; Fulconis, J.; Duligall, J.; Wadsworth, W. J.; Russell, P. St. J. {\em Opt. Express } {\bf 2005} {\em 13}, 534-544. 

\bibitem[3]{fan05} Fan, J.; Migdall, A. {\em Opt. Express} {\bf 2005} {\em 13}, 5777-5782. 

\bibitem[4]{uren05} U'Ren, A. B.; Silberhorn, Ch.; Erdmann, R.; Banaszek, K.; Grice, W. P.; Walmsley, I. A.; Raymer, M. G. {\em Las. Phys} {\bf 2005}, {\em 15}, 146--161.

\bibitem[5]{zhang07} Zhang, L.; U'Ren, A. B.; Erdmann, R.; O'Donnell, K. A.; Silberhorn, Ch.; Banaszek, K.; Walmsley, I. A. {\em J. Mod. Opt.} {\bf 2007} {\em 54}, 707--719.

\bibitem[6]{odonnell07} O'Donnell, K. A.; U'Ren, A. B. {\em Opt. Lett.} {\bf 2007} {\em 32}, 817--819.

\bibitem[7]{dayan05} Dayan, B.; Pe'er, A.; Friesem, A. A.; Silberberg, Y. {\em Phys. Rev. Lett.} {\bf 2004} {\em 93}, 023005.

\bibitem[8]{molotkov98} Molotkov, S. N. {\em JETP Lett.} {\bf 1998} {\em 68}, 263--270.

\bibitem[9]{nasr03} Nasr, M. B.; Saleh, B. E. A.; Sergienko, A. V.; Teich, M. C. et al. {\em Phys. Rev. Lett.} {\bf 2003} {\em 91}, 083601.

\bibitem[10]{grice01} Grice, W. P.; U'Ren, A. B.; Walmsley, I. A. {\em Phys. Rev. A} {\bf 2001} {\em 64}, 063815.

\bibitem[11]{mosley08} Mosley, P. J.; Lundeen, J. S.; Smith, B. J.; Wasylczyk, P.; U'Ren, A. B.; Silberhorn, Ch.; Walmsley, I. A. Phys. Rev. Lett. 2008, 100, 133601.

\bibitem[12]{garay07} Garay-Palmett, K.; McGuinness, H. J.; Cohen, O.; Lundeen, J. S.; Rangel-Rojo, R.; U'Ren, A. B.; Raymer, M. G.; McKinstrie, C. J.; Radic, S.; Walmsley, I. A.   {\em Opt. Express} {\bf 2007}, {\em 15}, 14870--14886. 

\bibitem[13]{agrawal2007} Agrawal, G. P. {\em Nonlinear Fiber Optics 4th Ed.}; Elsevier:2007.

\bibitem[14]{wadsworth02} Wadsworth, W. J.; Ortigosa-Blanch, A.; Knight, J. C.; Birks, T. A.; Man, T. -P. M.; Russell, P. S. J. {\em J. Opt. Soc. Am. B} {\bf 2002}, {\em 19}, 2148--2155. 

\bibitem[15]{Brambilla05} Brambilla, G.; Koizumi, F.; Feng, X.; Richardson, D. J. Electron. Lett. 2005, 41, 400-402. 

\bibitem[16]{Foster08} Foster, M. A.; Turner, A. C.; Lipson, M.; Gaeta, A. L. {\em Opt. Express} {\bf 2008}, {\em 16}, 1300--1320. 

\bibitem[17]{radic03} Radic, S.; McKinstrie, C. J.; Jopson, R. M.; Centanni, J. C.; Lin, Q.; Agrawal, G. P. {\em Electron. Lett.} {\bf 2003}, {\em 39}, 838--839. 

\bibitem[18]{brainis05} Brainis, E.; Amans, D.; Massar, S. {\em Phys. Rev. A} {\bf 2005}, {\em 71}, 023808.

\bibitem[19]{garay08} Garay-Palmett, K.; U'Ren, A. B.; Rangel-Rojo, R.; Evans, R.; Camacho-L\'opez, S. {\em To appear in Phys. Rev. A}.

\bibitem[20]{rohde} Rohde, P. P.; Mauerer, W.; Silberhorn, Ch. {\em New J. Phys} {\bf 2007}, {\em 9}, 91. 

\bibitem[21]{Tong06} Tong, L.; Hu, L.; Zhang, J.; Qiu, J.; Yang, Q.; Lou, J.; Shen, Y.; He, J.; Ye, Z. {\em Opt. Express} {\bf 2006}, {\em 14}, 82--87. 

\bibitem[22]{Ebendorff04} Ebendorff-Heidepriem, H.; Petropoulos, P.; Asimakis, S.; Finazzi, V.; Moore, R.; Frampton, K.; Koizumi, F.; Richardson, D.; Monro, T. {\em Opt. Express} {\bf 2004}, {\em 12}, 5082--5087. 

\bibitem[23]{Gopinath} Gopinath, J.; Shen, H.; Sotobayashi, H.; Ippen, E.; Hasegawa, T.; Nagashima, T.; Sugimoto, N. Opt. Express. 2004, 12, 5697-5702. 


\end{thebibliography}
\end{document}